\documentclass[]{spie}  

\usepackage{amsmath,amsfonts,amssymb}
\usepackage{graphicx}
\graphicspath{ {./figures/} }
\usepackage[colorlinks=true, allcolors=blue]{hyperref}
\usepackage{float}
\usepackage{gensymb}
\usepackage{textcomp}

\title{Stray and Scattered Light Considerations in a Non-contiguous Array of Commercial CMOS Sensors in a Space Mission}

\author[a]{Maggie Y. Kautz}
\author[a]{Douglas Kelly}
\author[a,b]{Heejoo Choi}
\author[b]{Young Sik Kim}
\author[a]{Fernando Coronado}
\author[a]{Cameron C. Ard}
\author[a]{Patrick Ingraham}
\author[a,b]{Daewook Kim}
\author[a]{Ewan S. Douglas}

\affil[a]{Steward Observatory, University of Arizona, 933 N Cherry Ave, Tucson, AZ, 85719}
\affil[b]{James C. Wyant College of Optical Sciences, University of Arizona, 1630 E University Blvd, Tucson, AZ, 85721}

\authorinfo{Further author information: (Send correspondence to M.K.)\\M.K.: E-mail: maggiekautz@arizona.edu}

\begin{document} 
\maketitle

\begin{abstract}
Recent advances in CMOS technology have potential to significantly increase the performance, at low-cost, of an astronomical space telescope. Arrays of sensors in space missions are typically contiguous and act as a monolithic detector. A non-contiguous array, with gaps between individual commercial CMOS detectors, offers potential cost and schedule benefits but poses a unique challenge for stray/scattered light mitigation due to complexities in the optomechanics. For example, if the array of detectors is being fed a large field of view, then each detector will have a different angle of incidence. Any individual bandpass filters need to be held perpendicular to the incoming beam so as not to create variances of central wavelength transmission from detector to detector. It naturally follows that the optical design can force filter ghosts to fall between detectors. When dealing with well-focused, high-intensity beams, first and second order stray light path analyses must be conducted to determine scattered light from glints off of individual optics/opto-mechanics or detector specific vane structures. More mechanical structures are necessary for imaging with non-contiguous arrays, all of which have potential to increase scattered light. This proceeding will document various stray light mitigation strategies for a non-contiguous array of sensors in a space telescope. 
\end{abstract}


\section{INTRODUCTION}
\label{sec:intro}

In a wide-field context camera, there needs to be pixel coverage of the entire field-of-view (FOV) of the optical system\cite{kim_compact_2023, kim_large_2025}. The primary benefits of using a non-contiguous array of commercial CMOS detectors, as opposed to a custom singular detector design, are cost and schedule\cite{douglas_approaches_2023}. A much higher pixel resolution can be achieved at a significantly reduced cost compared to a custom system. Modularity within the whole detector system allows for changes to the upstream optical system and its subsequent FOV. Tolerances, such as those for alignment, for off and on-axis detectors don't necessarily need to be the same. Additionally, different detectors can perform different tasks. For example, some may be utilized for guiding while others could be used for wavefront sensing. However, there are inherent risks to this set-up. Multiple sensors may require distinct narrow-band filtering systems with distinct mounting structures. This introduces ghosting from detector to detector in addition to ghosting on a singular detector. More mechanical structures also lead to more stray light opportunities. Degraded optical performance of space missions has been attributed to stray light glints off of optics and structural supports in traditional detector schematics\cite{brendel_balloon-borne_2022}. This proceeding will discuss unique stray light mitigation strategies for a non-contiguous array of commercial sensors in a space mission.

\section{Stray and Scattered Light Mitigation}
\label{sec:stray}

\subsection{Optomechanical Complexities}
\begin{figure} [H]
\begin{center}
\begin{tabular}{c} 
\includegraphics[width=0.75\textwidth]{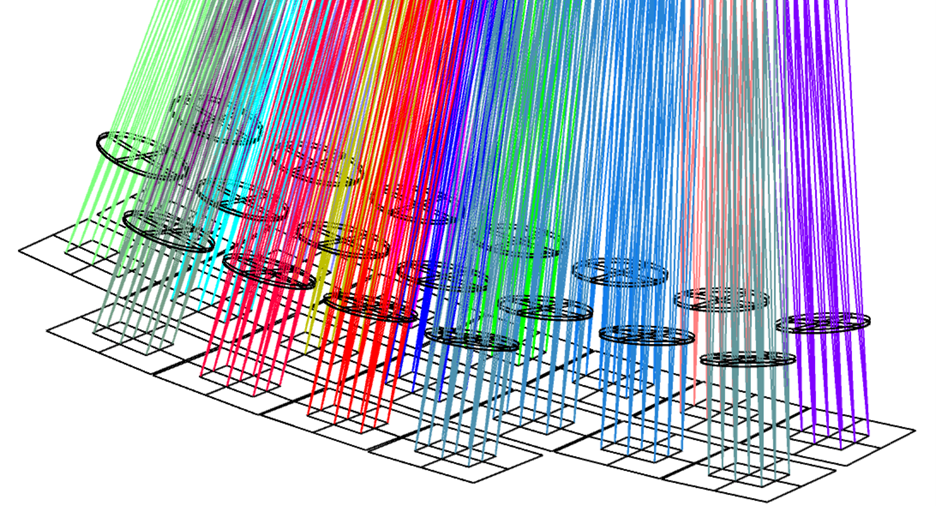}
\end{tabular}
\end{center}
\caption{Zemax rendering rays spread out over 18 individual adjacent CMOS detectors.}
\label{fig:lazuli_rays_close} 
    \end{figure}

\begin{figure} [H]
\begin{center}
\begin{tabular}{c} 
\includegraphics[width=0.75\textwidth]{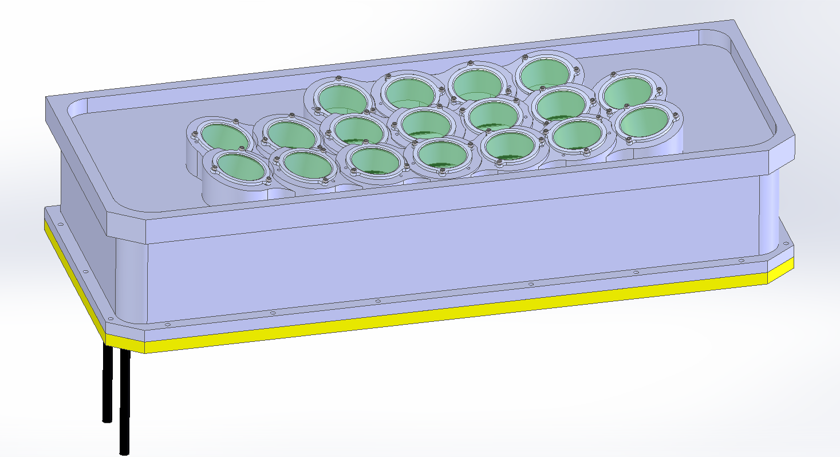}
\end{tabular}
\end{center}
\caption{CAD rendering of optomechanical structure for an array of commercial sensors.}
\label{fig:lazuli_iso_view} 
    \end{figure}

An array of commercial sensors (Figure~\ref{fig:lazuli_rays_close}) requires a complex optomechanical mounting structure, such as the one shown in Figure~\ref{fig:lazuli_iso_view}. While a benefit to a modular system like this is that different detectors can be performing different tasks, some detectors may need to utilize the same bandpasses for imaging. Since the array of detectors is being fed a large field of view, each sensor will have a unique angle of incidence. Any individual bandpass interference filter needs to be held perpendicular to the incoming beam so as not to create variances of central wavelength transmission from detector to detector.

\begin{figure} [H]
\begin{center}
\begin{tabular}{c} 
\includegraphics[width=0.7\textwidth]{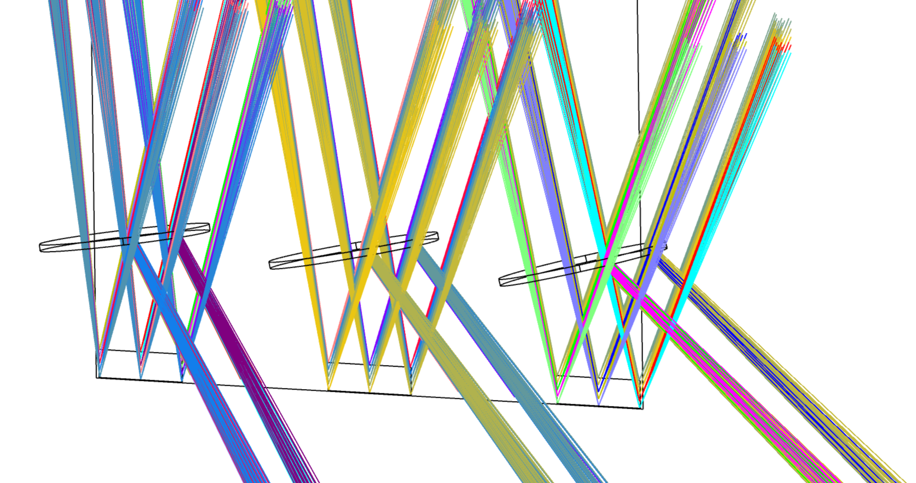}
\end{tabular}
\end{center}
\caption{Zemax rendering showing filter optics placed so there is no ghosting back onto the detector or onto the adjacent detector.}
\label{fig:ghost} 
    \end{figure}

When employing various upstream optics, such as this array of bandpass filters, ghosting needs to be taken into account. The filters need to be placed high enough so that the filters do not allow light reflected up from the detector to be reflected back down from the top or bottom of the optic back onto the detector. However, the filters cannot be placed so high that ghosts can appear on an adjacent detector (Figure~\ref{fig:ghost}).

\begin{figure} [H]
\begin{center}
\begin{tabular}{c} 
\includegraphics[width=0.75\textwidth]{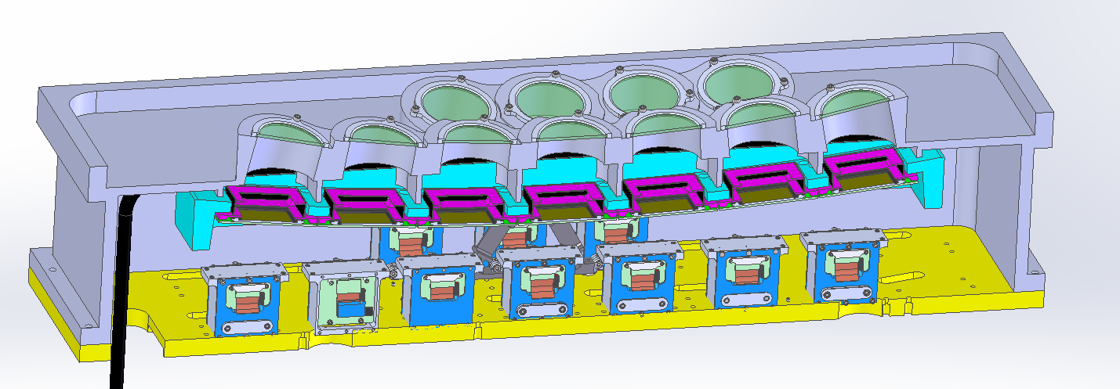}
\end{tabular}
\end{center}
\caption{Section view of above optomechanical structure showing stray light mitigation vanes and filter housings.}
\label{fig:sec_view} 
    \end{figure}

\begin{figure} [H]
\begin{center}
\begin{tabular}{c} 
\includegraphics[width=0.8\textwidth]{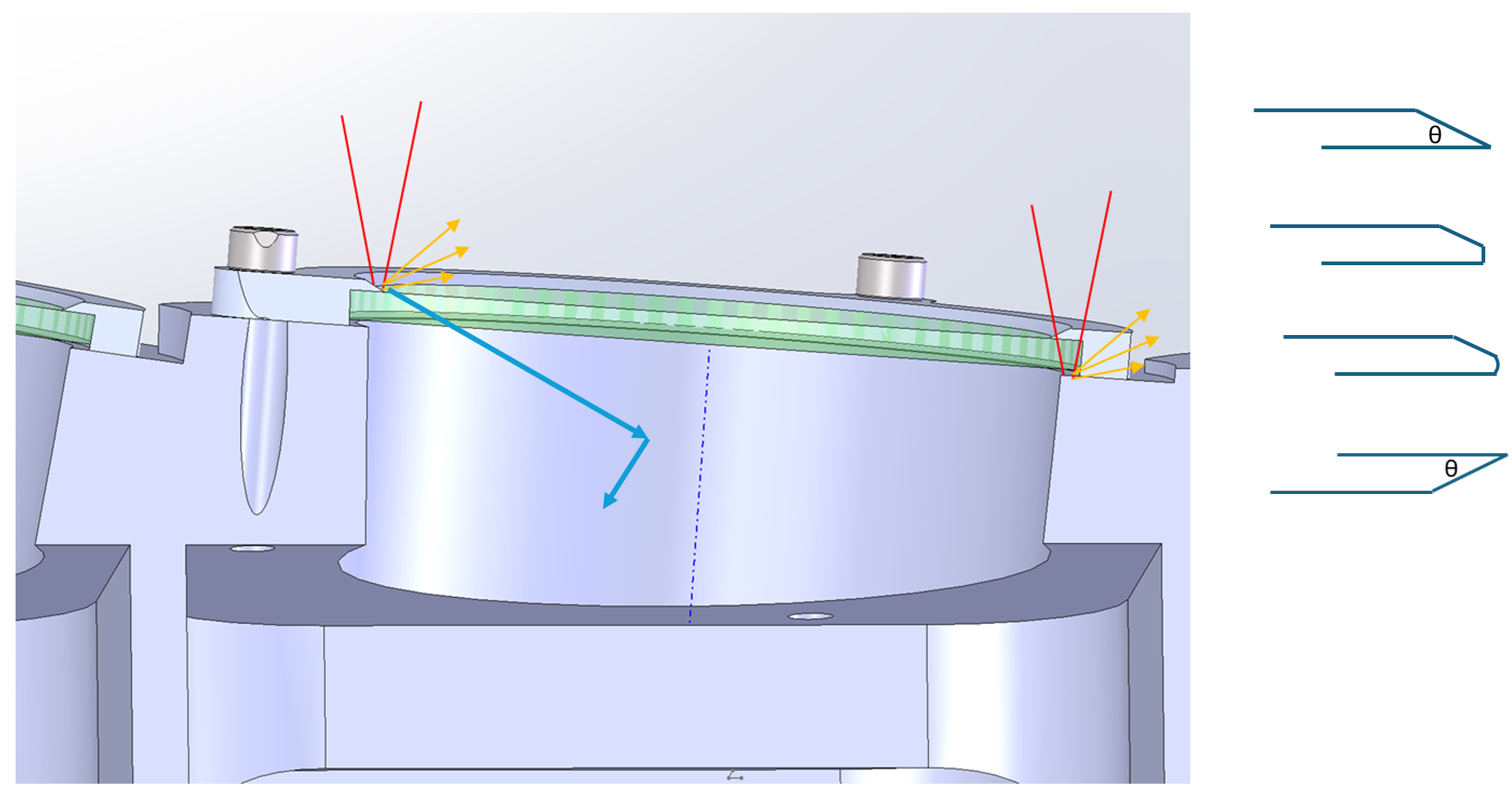}
\end{tabular}
\end{center}
\caption{This is a section view of a singular filter's mounting structure to show specific edges where stray light should be analyzed. The red lines indicate incident light rays, the yellow arrows indicate scattered light, and the blue arrows indicate scattered light following a path towards the detector under the filter. Examples of various edge types are shown to the right.}
\label{fig:rr_sec_view} 
    \end{figure}

\begin{figure} [H]
\begin{center}
\begin{tabular}{c} 
\includegraphics[width=0.75\textwidth]{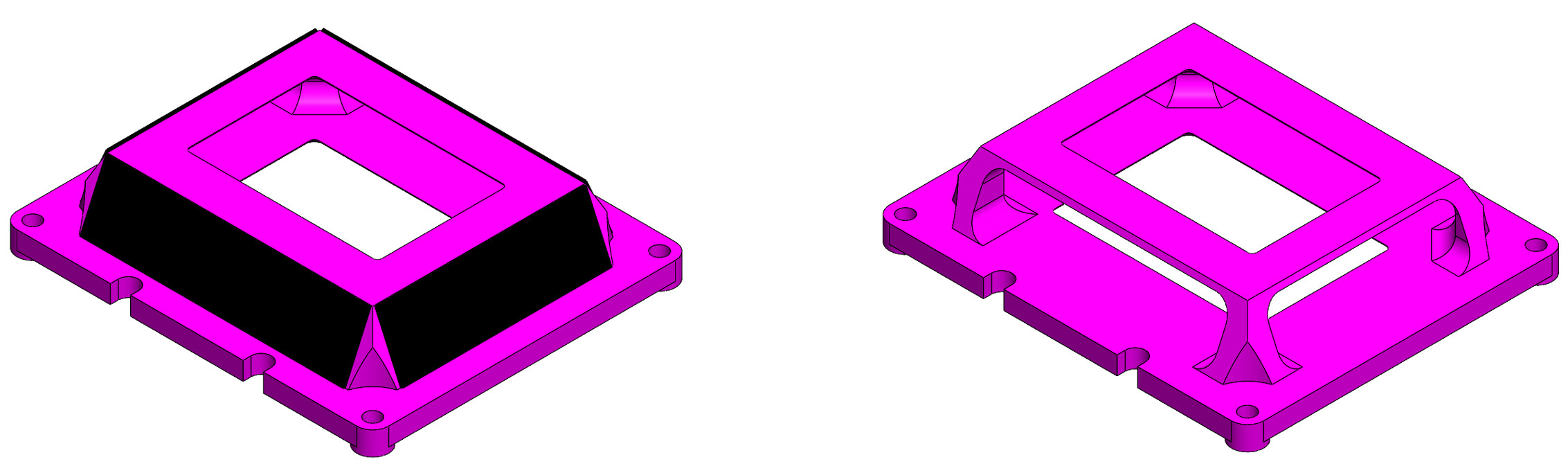}
\end{tabular}
\end{center}
\caption{On the left there is a two tier vane structure enclosed within walls to stop light scatter between detectors. On the right is an open version of the same vane structure to better display the design.}
\label{fig:vanes} 
    \end{figure}

Figure~\ref{fig:sec_view} shows a section view of the structure shown in Figure~\ref{fig:lazuli_iso_view}. Multiple levels of baffling, specifically the minimum required, are implemented. Immediately below an individual filter optic, there is a cylindrical baffle extending downwards towards the detector (Figure~\ref{fig:rr_sec_view}). This, combined with an extended wall structure, prevents stray light bounces between detectors. A two-tier vane structure closer to the detector is utilized for blocking higher intensity, closer-to-focus, stray light (Figure~\ref{fig:vanes}).

The edges of each structure with an optical aperture need be evaluated for how light will scatter when incident upon them. The angle/orientation of the bevel/chamfer or the tolerance of a knife edge tip are all parameters to be studied in a stray light analysis. For example, round edges, perhaps from paint application, may yield worse stray light performance than a knife edge\cite{kim_stray_2022}. While a knife edge may be ideal, cost/manufacturability needs to be considered since this space mission is already employing commercial detectors for cost/schedule benefits. This means that various tolerances of a knife edge should be considered when modeling how light may travel downward towards a detector. That is to say the knife edge may become more blunt (edge image second from top in Figure~\ref{fig:rr_sec_view}).

Closer to the focus on the detector, light will be more concentrated. Tightly toleranced, black painted vanes can be utilized to block this higher intensity incident light (Figures~\ref{fig:vanes}).

\subsection{Tolerancing}

\begin{figure} [H]
\begin{center}
\begin{tabular}{c} 
\includegraphics[width=0.75\textwidth]{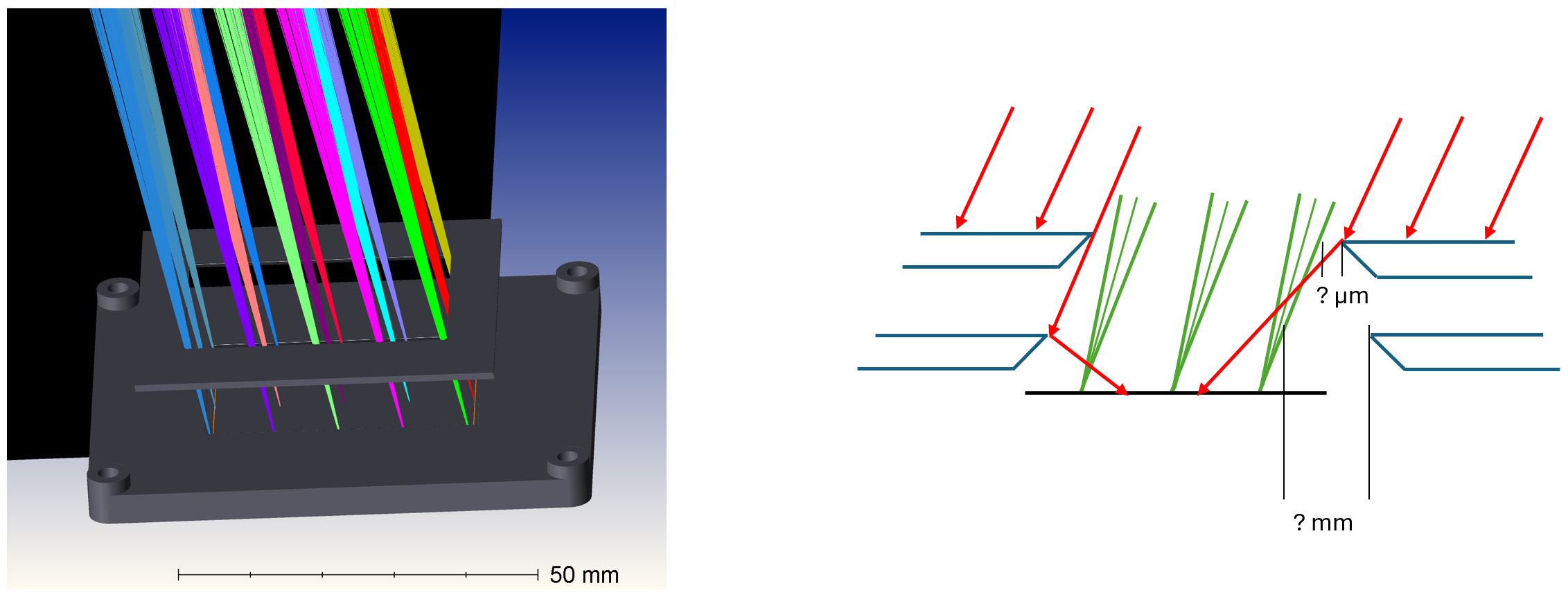}
\end{tabular}
\end{center}
\caption{The left image shows multiple fields passing through essentially the same type of two-tier vane structure as shown in Figure~\ref{fig:vanes}. The right image highlights the tolerances on these vanes, ie how close they are allowed to be to the science beam. The question mark indicates the manufacturing tolerance on the vane which dictates how close it can be to the outer marginal ray of the outer science field.}
\label{fig:manu_tol} 
    \end{figure}

Manufacturing and alignment tolerances are critical parts of stray light mitigation. Figure~\ref{fig:manu_tol} shows a two-tier vane structure, similar to the one shown in Figure~\ref{fig:vanes}, with a field of beams filling the aperture and incident on an individual detector within an array. How close the edges of the vane can be to the outer marginal rays without causing vignetting or allowing out-of-field stray light to hit the detector is a tolerancing problem.

\begin{figure} [H]
\begin{center}
\begin{tabular}{c} 
\includegraphics[width=0.75\textwidth]{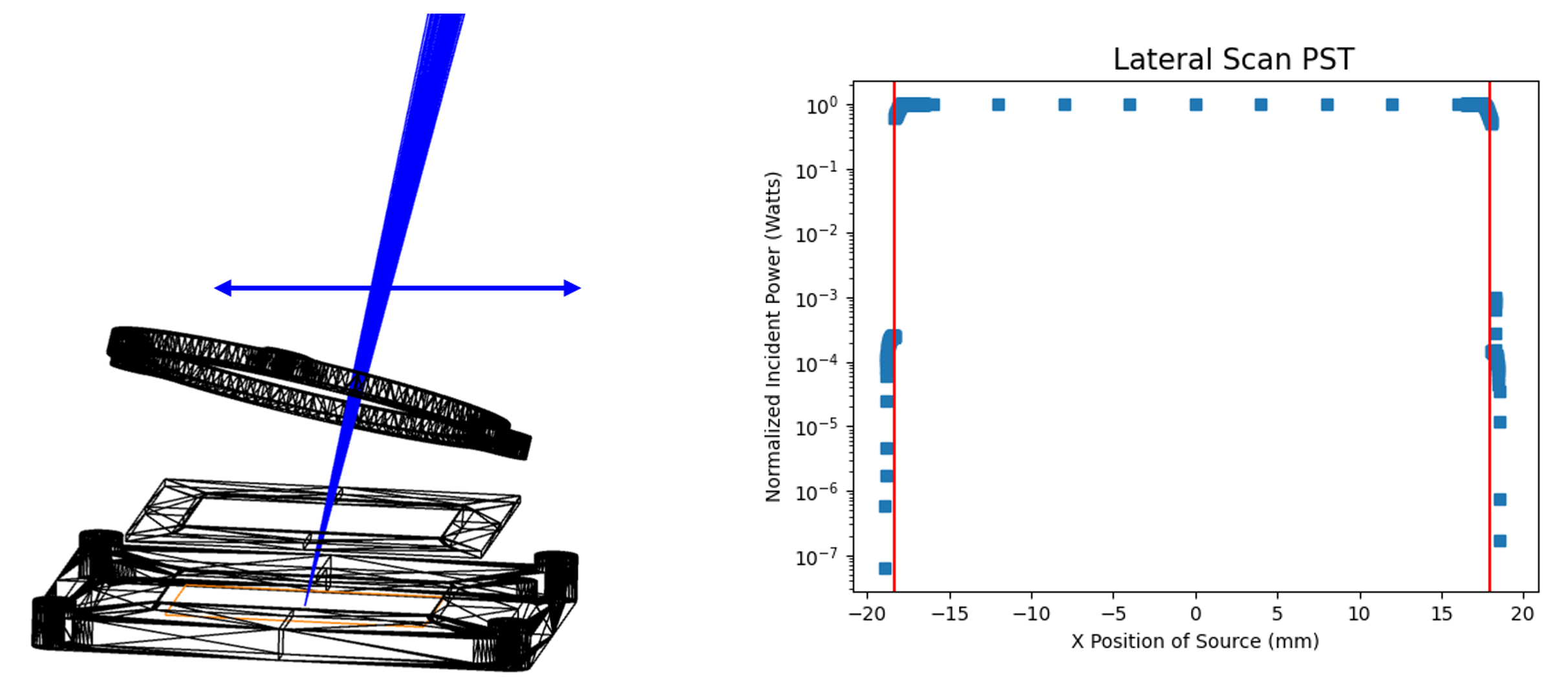}
\end{tabular}
\end{center}
\caption{The left Zemax OpticStudio\textsuperscript{\textregistered} image shows the two-tier vans structure and filter housing that were analyzed in a lateral scan PST. The right image shows the results of the lateral scan PST. Stray light drops down to a level of 10\textsuperscript{-4} very quickly when the upper vane is toleranced at 100 $\mu$m from the outer marginal ray.}
\label{fig:pst} 
    \end{figure}

A point source transmission function (PST) expresses the stray light level at a detector from a point source at infinity (effectively collimated) illuminating an optical system’s entrance pupil and passing through the system. It can be defined as a point source’s integrated irradiance at the final focal plane (detector) relative to its incident irradiance at the entrance pupil, at an angle relative to the center of the FOV\cite{king_radiometry_2024,freniere_soar_1990}. A ``lateral scan" PST is a take on a normal PST where a focusing beam is traversed across the entrance aperture of an optical system, in this case the upper vane opening, and the corresponding stray light levels are measured (Figure~\ref{fig:pst}). When the beam traverses over the edge it will vignette and scatter. The goal of the vane design and tolerance trade space is to minimize vignetting within the science field while rapidly dropping off scattered light out of the science field.

\begin{figure} [H]
\begin{center}
\begin{tabular}{c} 
\includegraphics[width=0.5\textwidth]{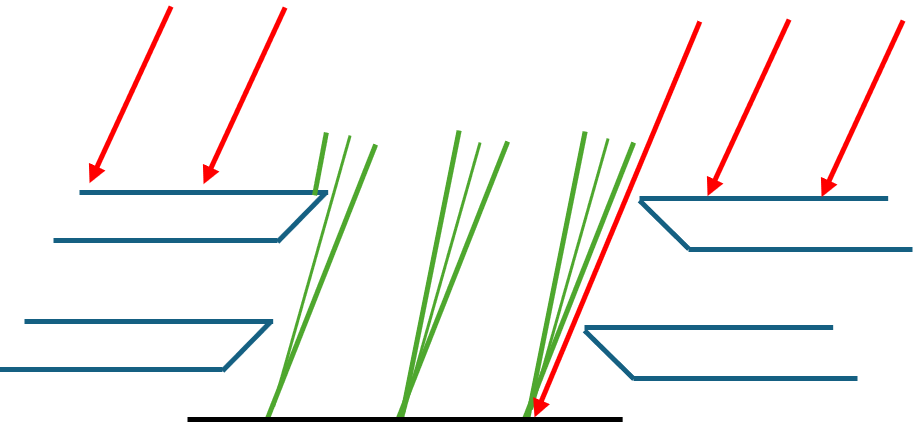}
\end{tabular}
\end{center}
\caption{When the upper vane structure is misaligned it can allow vignetting, shown on the left, and stray light, shown on the right.}
\label{fig:align_tol} 
    \end{figure}

Manufacturing tolerances are no good if the structures are not aligned well. Figure~\ref{fig:align_tol} shows that if the upper vane is not aligned properly, stray light and vignetting can be introduced. A next step in this project is to define an alignment procedure for the optomechanical structure in order to meet alignment tolerances that fall within manufacturing tolerances.

\section{CONCLUSION}

A non-contiguous array of individual commercial CMOS detectors offers significant cost and schedule benefits to a space mission but introduces stray/scattered light due to the complexities in the optomechanics. Baffling and vane structures can be implemented to mitigate stray light. It is necessary to analyze edge structures and upstream optics for scattered light and ghosting purposes. The overall goal of the baffle/vane design and tolerance trade space is to minimize vignetting within the science field while stopping scattered light out of the science field from reaching the detector.
 
\acknowledgments  
 
Portions of this research were supported by funding from the Technology Research Initiative Fund (TRIF) of the Arizona Board of Regents and by generous philanthropic donations to the Steward Observatory of the College of Science at the University of Arizona.

\bibliography{SPIE_2025_Proc} 
\bibliographystyle{spiebib} 

\end{document}